# Assessment of the environmental impacts of the Cherenkov Telescope Array Mid-Sized Telescope

Gabrielle dos Santos Ilha[1], Marianne Boix[2], Jürgen Knödlseder[1], Philippe Garnier[1], Ludovic Montastruc[2], Pierre Jean[1], Giovanni Pareschi[3], Alexander Steiner[4], François Toussenel[5]

1 Institut de Recherche en Astrophysique et Planétologie, Université de Toulouse, CNRS, CNES, UPS, 31000 Toulouse, France

2 Laboratoire de Génie Chimique, U.M.R. 5503 CNRS/INP/UPS, Université de Toulouse, 4, Allée Emile Monso, 31432 Toulouse Cedex 4, France

3 INAF, Osservatorio Astronomico di Brera, Via E. Bianchi 46, 23807 Merate (Lc), Italy

4 Deutsches Elektronen-Synchrotron DESY, Platanenallee 6, 15738 Zeuthen, Germany

5 Sorbonne Université, CNRS/IN2P3, Laboratoire de Physique Nucléaire et de Hautes Energies, LPNHE, 4 place Jussieu, 75005 Paris, France

**Astronomical observatories have been identified as substantial contributors to the carbon footprint of astrophysical research. Being part of the collaboration that currently develops the Medium-Sized Telescope (MST) of the Cherenkov Telescope Array, a ground-based observatory for very-high-energy gamma rays that will comprise 64 telescopes deployed on two sites, we assessed the environmental impacts of one MST on the Northern site by means of a Life Cycle Assessment. We identified resource use and climate change as the most significant impacts, being driven by telescope manufacturing and energy consumption during operations. We estimate life cycle greenhouse gas emissions of 2,660 ± 274 $tCO_2$ equivalent for the telescope, 44% of which arise from construction, 1% from on-site assembly and commissioning, and 55% from operations over 30 years. Environmental impacts can be reduced by using renewable energies during construction and operations, use of less electronic components and metal casting, and use of recycled materials. We propose complementing project requirements with environmental budgets as an effective measure for impact management and reductions.**

Recent research in the context of the planetary boundaries framework has evidenced that human activities heavily perturb the Earth system as a whole, pushing it well outside of the safe operating space for humanity [1]. Symptoms of this perturbation include, but are not limited to, global warming [2], biodiversity loss [3], depletion of natural resources [4], and pollution of air, soil and water [5]. Returning back into safe operating space hence implies reducing anthropogenic impacts on the Earth system, taking a systemic approach that avoids shifting impacts from one area of concern to another.

While scientific research contributes significantly to our understanding of the world in which we are living, there is increasing recognition that research activities, as all human activities, have an impact that eventually contributes to the deterioration of the environment [6]. The astronomical research community is among the communities that are particularly active in this introspection, and estimates of greenhouse gas (GHG) emissions related to astronomical research activities are becoming part of the scientific production of the field [7]. Primary sources of GHG emissions include air travelling, purchase of goods and services, and use of research infrastructures, which are ground-based or space-based observatories. It turns out that the latter are the dominant source of GHG emissions in the field, with an estimated average contribution of 36.6 ± 14.0 $tCO_2$ equivalent ($CO_2$eq) per astronomer [8]. Yet, little is known so far about the origin of these emissions, and hence about possible lever-arms for emission reductions. Furthermore, environmental impacts other than GHG emissions have so far attracted little attention.

This situation is about to change, with first Life Cycle Assessments (LCAs) of astronomical research infrastructures appearing in the literature. For example, Kruithof et al. [9] estimated the lifecycle carbon footprint of the Low Frequency Array (LOFAR), finding that 70% of the carbon footprint arises in the operations phase. Vargas-Ibáñez et al. [10] performed a simplified LCA for the Giant Radio Array for Neutrino Detection prototype (GRANDProto300), identifying the construction of the antenna structures and batteries as the processes with the highest environmental impacts, with resource use, ionising radiation, climate change and acidification being the most important impact categories. Viole et al. [11] conducted a comparative LCA to study various renewable-based energy systems to power the AtLAST telescope in the Atacama Desert, identifying high-renewable systems with some fossil fuel generation as the best compromise between power reliability, GHG emissions, mineral resource depletion and water use. Barret et al. [12] performed a LCA to estimate the environmental impacts associated with the development of the X-IFU instrument aboard ESA's Athena X-ray satellite, identifying resource extraction for instrument model manufacturing and energy consumption for instrument testing in clean rooms as most important sources of environmental impacts.

To go beyond these initial works, we assessed for the first time the environmental impacts of a ground-based astronomical telescope over its entire life cycle using standard LCA analysis methodology and tools (see Methods for details). Our study presents a substantial effort in terms of modelling detail and accuracy that provides a comprehensive view of the environmental impacts that are generated by the manufacturing and operation of astrophysical research infrastructures. We performed our assessment for one Mid-Sized Telescope (MST) of the Cherenkov Telescope Array Observatory (CTAO) that will be deployed on the Observatory of the Roque de los Muchachos (ORM) on the Canary island of La Palma, Spain, where CTAO's Northern site is located. CTAO will be the largest and most advanced ground-based observatory for detecting gamma rays produced by cosmic particle accelerators such as supermassive black holes and supernovae, covering an energy range from 20 GeV to 300 TeV. CTAO will involve two arrays of telescopes, one in the Southern Hemisphere, in the Atacama Desert in Chile, and the other in the Northern Hemisphere at ORM. A total of 64 telescopes will be built, comprising 37 Small-Sized Telescopes (SSTs) and 14 MSTs in the Southern Hemisphere, and 9 MSTs and 4 Large-Sized Telescopes (LSTs) in the Northern Hemisphere [13].

The MST-North (MSTN) system that we studied consists of 464 tonnes of concrete foundations, 82 tonnes of telescope structure (MST-STR), and a NectarCAM camera that weighs 2.2 tonnes (tons refer to metric tons throughout the paper). MST-STR is provided by an international consortium of research institutes led by the Deutsches Elektronen-Synchrotron (DESY), located in Zeuthen, Germany, while NectarCAM is provided by an international consortium of research institutes led by the Institut de Recherche sur les lois Fondamentales de l'Univers (IRFU), located in Saclay, France (see 'Overview over the Mid-Sized Telescope' in Supplementary Information). We modelled the MST-STR and NectarCAM units independently, including an estimated amount of 5% of spare parts that will be used for maintenance during 30 years of operations, which is the required lifetime of the system. After 15 years of operations, it is expected that the camera will be upgraded by replacing either parts or the entire camera by a new system. Since no plans or specifications exist so far for the second generation camera, we assumed that its environmental impacts will be identical to those of the NectarCAM. The life cycle therefore comprises the construction of one MST-STR and two NectarCAMs, and operations of a system composed of MST-STR and NectarCAM during 30 years. Operations include electricity consumed for operating the telescope, as well as labour and associated travelling required for the maintenance of the system. We furthermore assumed that the telescope and camera are constructed now, with delivery to site, commissioning and start of operations immediately after that.

The LCA was performed using the SimaPro software version 9.5.0 in combination with the Life Cycle Inventory (LCI) database ecoinvent version 3.9.1. The functional unit of our assessment was "One MST constructed, operated and maintained for a lifetime of 30 years on the CTAO North site in La Palma, Spain, to observe air Cherenkov showers". The system boundaries for the LCA are depicted in Extended Data Fig. 1. MSTN specific LCI data were collected from project documentation, through expert interviews, but also through direct measurements. For each MSTN-specific data item, representative LCI processes were identified in the ecoinvent database, tracking data quality and representativeness of the ecoinvent processes for an assessment of uncertainties in the environmental impacts (see Methods). We used the Environmental Footprint 3.1 method for our analysis.

# Results

The life cycle impacts of MSTN that we computed from the LCI and their 95% confidence level uncertainties that we determined using Monte Carlo simulations are summarised in Table 1 for all environmental impact categories. Relative contributions of the construction of foundations, MST-STR and NectarCAM, as well as telescope deployment and operations are also given.

The relative contributions of the different activities to the life cycle impact of one MSTN are illustrated in Figure 1. Telescope construction dominates 10 out of 16 impact categories, while telescope operations dominate the 6 other impact categories. Impacts of deployment are negligible. Owing to the fact that almost 90% of the electricity of La Palma is currently produced by diesel combustion, electricity consumption during operations has a particularly large climate change impact. Note that we have assumed in our analysis that the electricity system in La Palma will stay as it currently is over next 30 years, which may be considered conservative in view of stated goals to move to 100% renewable energy on the island by 2030, but which may actually be realistic as there currently exist no concrete implementation plans for such a transformation [14]. Nevertheless, we have alternatively modelled two possible future renewable energy systems in 'Alternative energy systems for La Palma' in the Supplementary Information, with resulting MSTN life cycle impacts shown in Extended Data Figs. 2 and 3, demonstrating that if a renewable energy system would be implemented on La Palma, the environmental impact of MSTN operations would be drastically reduced.

**Table 1. Environmental life cycle impacts of one MSTN.** The table provides the total impacts for all categories as well as the relative contributions of foundations, MST-STR and NectarCAM construction, telescope deployment on La Palma, and operations for 30 years. Quoted uncertainties were determined using Monte Carlo simulations and are for 95% confidence levels. Note that freshwater ecotoxicity, human toxicity, ionising radiation and water use have relatively large uncertainties compared to the values of the impacts which is an inherent feature of the ecoinvent database that was used in the study.

| Environmental impact category | Unit | Total | Foundations | MST-STR | NectarCAM | Deployment | Operations |
|---|---|---|---|---|---|---|---|
| Acidification | mol H+ eq | 16.9 ± 2.4 | 2.0% | 31% | 9% | 0.5% | 57% |
| Climate change | t $CO_2$ eq | 2,660 ± 274 | 3.5% | 33% | 8% | 0.7% | 55% |
| Ecotoxicity, freshwater | MCTUe | 17.2 ± 352 | 1.9% | 27% | 25% | 0.7% | 46% |
| Particulate matter | disease inc. | 0.148 ± 0.045 | 3.2% | 39% | 12% | 0.4% | 46% |
| Eutrophication, marine | kg N eq | 3,987 ± 940 | 2.4% | 31% | 8% | 0.6% | 58% |
| Eutrophication, freshwater | kg P eq | 610 ± 446 | 3.2% | 59% | 33% | 0.1% | 5% |
| Eutrophication, terrestrial | kmol N eq | 41.4 ± 10.3 | 2.6% | 30% | 8% | 0.6% | 59% |
| Human toxicity, cancer | CTUh | 0.003 ± 0.111 | 5.9% | 71% | 7% | 0.2% | 16% |
| Human toxicity, non-cancer | CTUh | 0.027 ± 25.8 | 3.1% | 50% | 31% | 0.5% | 16% |
| Ionising radiation | MBq U-235 eq | 211 ± 509 | 0.9% | 35% | 59% | 0.1% | 4% |
| Land use | kPt | 7.7 ± 2.0 | 5.0% | 48% | 28% | 0.4% | 18% |
| Ozone depletion | kg CFC11 eq | 0.057 ± 0.015 | 1.6% | 30% | 18% | 0.6% | 49% |
| Photochemical ozone formation | t NMVOC eq | 13.8 ± 3.1 | 2.6% | 30% | 7% | 0.7% | 59% |
| Resource use, fossils | TJ | 33.9 ± 7.1 | 2.2% | 32% | 14% | 0.7% | 51% |
| Resource use, mineral & metals | kg Sb eq | 63.6 ± 33.5 | 0.5% | 19% | 79% | 0.1% | 2% |
| Water use | $Mm^3$ depriv. | 0.321 ± 118 | 8.8% | 39% | 11% | 0.3% | 41% |

MST-STR construction dominates the freshwater eutrophication and human toxicities impact categories. Use of stainless steel contributes to the human toxicity impacts, with cancer impacts originating from landfill of residual materials arising from electric arc furnace slag, while non-cancer impacts originating from mining of ferronickel ores, the melting of metals, and the production of electricity and heat. Treatment of spoils by surface landfill also contributes to freshwater eutrophication. Even more significant impacts are arising from steel casting, a process that is used for the head and yokes of the positioner of the telescope. Steel casting consumes large amounts of energy, requires adding an extra amount of stainless steel, and involves treatment of spoils of lignite and hard coal mining activities by surface landfill. They dominate the land use impacts, owing to furnace heat production, electricity production from oil and coal, landfill of residuals, and mining activities. Steel casting also contributes significantly to freshwater eutrophication.

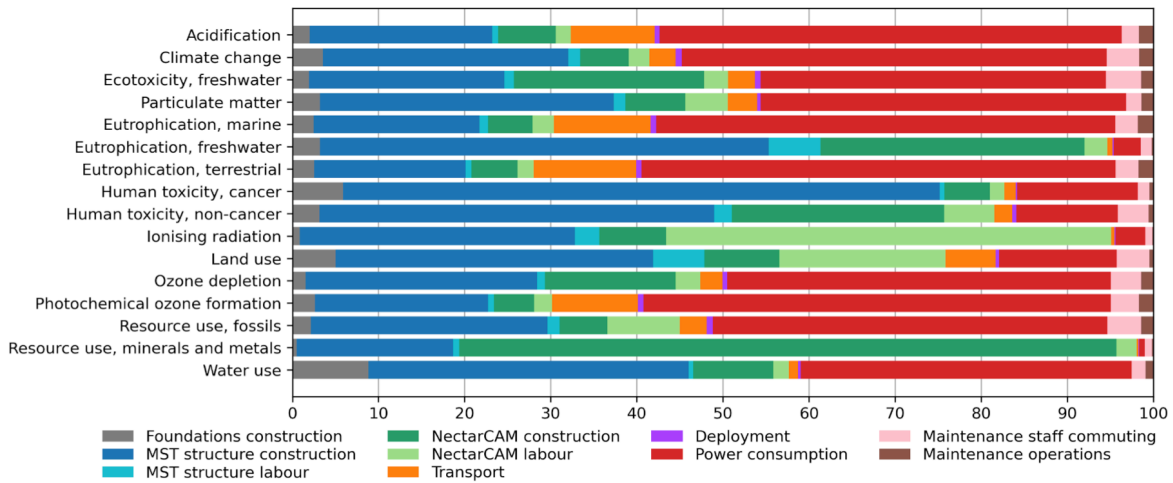

**Figure 1. Contribution of activities to the life cycle impacts of one MSTN.** Horizontal bars correspond to different environmental impact categories, the various activities are distinguished by different colours. All bars sum up to a total of 100%.

Mineral resource depletion and ionising radiation are particularly important for NectarCAM construction. The former is related to the use of gold in the electronics, mainly in integrated circuits, that are present on many electronic boards in the camera. The latter is related to electricity production in France, where most of the staff needed for NectarCAM production is located, owing to the large proportion of nuclear power in the French electricity mix.

The contribution of labour to the construction footprint of MST-STR is rather modest for all impact categories, while labour has an important impact for NectarCAM construction on some impact categories, including in particular ionising radiation, as well as land use, fossil resource use, and non-cancer human toxicity, all related to the power and heating of the office buildings that host the labour.

Transport is to a large extent dominated by MST-STR in all impact categories, which is plausible in view of the important weight of the structure (82 t) with respect to the camera (2.2 t). A substantial fraction of these impacts is related to delivery of the full structure to the site, with shipments by truck from DESY, Zeuthen (Germany) to Barcelona (Spain), and by sea freight from Barcelona to La Palma.

Deployment, which includes on-site assembly and commissioning, contributes only marginally to the environmental impacts, yet so far no field experience with MST commissioning exists, and hence this category may be underestimated.

Finally, maintenance contributes only moderately to the life cycle impacts of MSTN. Most contributions from maintenance are due to the commuting of maintenance staff by plane, assuming that maintenance staff flies from Barcelona to La Palma every 8 days. Depending on the precise residence of the maintenance staff, the actual impact of maintenance may differ. We also have not included off-site maintenance activities in our estimate, such as related to repairs of equipment, since no estimates are yet available on their frequency.

As climate change is one of the most salient environmental impact categories, we summarise in Table 2 our $CO_2$eq emission estimates for the different LCI categories. Assuming no change in the energy system, about 52% of the emissions will arise from electricity consumption during the 30 years of operations on La Palma, originating mainly from the combustion of fossil fuels. About 38% of the emissions originate from the production of mechanics and electronics as well as manufacturing processes, totalling to about 1 ktCO$_2$eq. With regards to the manufacturing category, a large part of the impact is related to the steel casting process, used in the head and yokes of the positioner, producing 15% of the emissions or a value of 400 tCO$_2$eq.

Table 2. Total CO$_2$eq life cycle emissions of one MSTN. Emissions are separated by LCI category.

| LCI category | tCO$_2$eq | Contribution |
|---|---|---|
| Electricity | 1,369 | 52% |
| Manufacturing | 573 | 22% |
| Mechanics | 297 | 11% |
| Electronics | 135 | 5% |
| Transport | 81 | 3% |
| Labour | 204 | 8% |
| Total | 2,660 | 100% |

To compare the different impact categories, we show in Figure 2 the normalised impacts, using the normalisation factors of Environmental Footprint 3.1 (adapted) method version 1.00, which correspond to the annual average per person impact. For example, a normalised value of 1000 corresponds to the annual average impact generated by 1000 people in the world.

The most significant normalised impact category is mineral and metal resource use, originating from the construction of the NectarCAM cameras. As already explained above, this impact arises from the use of gold in the camera electronics, primarily in integrated circuits that are employed for the amplification and digitisation of the signal and the triggering of the camera modules. NectarCAM construction contributes also importantly to freshwater eutrophication, which is arising from sulfidic tailings produced during gold mine operations.

The second most significant impact category is fossil resource use, related to energy generation on La Palma that currently originates to almost 90% from the combustion of diesel. Diesel combustion also has significant impacts in many other categories, including photochemical ozone formation, climate change, air acidification, freshwater ecotoxicity, terrestrial and marine eutrophication, and particulate matter.

At the third place is a group formed by freshwater eutrophication, climate change and photochemical ozone formation with impacts between about 340 and 380. Freshwater eutrophication arises to 95% from the construction of the telescope, with important contributions from steel casting of the head and yokes of the positioner and the fabrication of printed wiring boards of the camera. Climate change impacts have a mixed origin, with 50% arising from powering the telescope during the 30 years of operations, 45% from constructing and deploying the telescope, as well as upgrading the camera, and 5% from maintenance activities. Photochemical ozone formation impacts show a comparable share, with 54% arising from powering the telescope, 41% from construction and deployment, and 5% from maintenance.

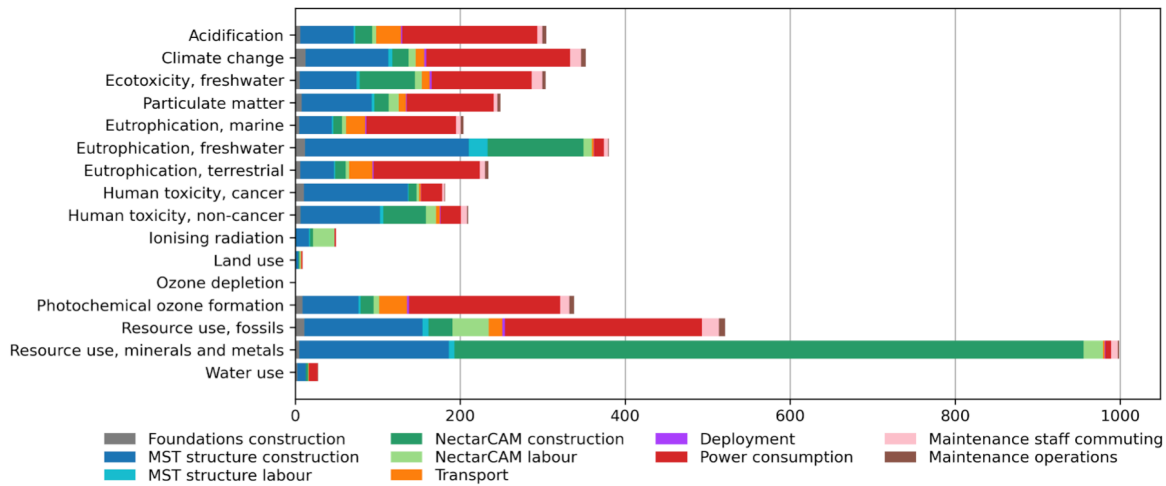

**Figure 2. Normalised life cycle impacts of one MSTN.** The values reflect integrated impacts from construction and 30 years of operations. They are given in units of the annual impacts of one average person in the world.

Acidification and freshwater ecotoxicity form a group with comparable impacts of about 300, while a further group with comparable impacts between about 230 and 250 is formed by particulate matter and terrestrial eutrophication. With impacts between about 180 and 210, marine eutrophication, cancer and non-cancer human toxicity are close to this group. Ionising radiation and water use have rather small normalised impacts, between about 30 and 50, while land use has an impact of about 10 and ozone depletion of about 1.

To allow for aggregation of the different impact categories, normalised results need to be weighted, and we show in Figure 3 the weighted impacts as derived using the Environmental Footprint 3.1 (adapted) method version 1.00 (see [22] for details on the determination of the weights). Weighted impacts depend on expert judgments on the importance of each impact category, and hence introduce some subjectivity in the analysis. According to this judgement, impacts for certain categories decrease with respect to the normalised impacts, such as photochemical ozone formation, freshwater ecotoxicity, marine, freshwater and terrestrial eutrophication, and human cancer and non-cancer toxicity, while others increase, in particular climate change. Weighted impacts are expressed in "points" (Pt), where one Pt corresponds to the annual environmental load of one average person in the world.

We aggregated the impacts in Figure 4 which shows the single score impacts for the different activities. Overall, MSTN construction has a larger contribution to the life cycle impact compared to operations. The most significant impact category for construction is the use of mineral and metal resources for NectarCAM, followed by climate change and fossil resource use. For operations, the most significant impact category is climate change, followed by fossil resource use and comparable contributions from acidification, particulate matter, and photochemical ozone formation.

We also note that the overall impacts of MST-STR and NectarCAM construction are comparable, despite the significant difference in their masses (82 t for the structure and 4.4 t for the cameras, considering that two cameras will be constructed during the life cycle of MSTN). This can be explained by the different impact categories involved, where impacts of MST-STR are primarily related to the energy consumption of manufacturing methods (metalworking and casting) and use of metals, while for NectarCAM the impact is primarily related to use of minerals and metals for electronic components.

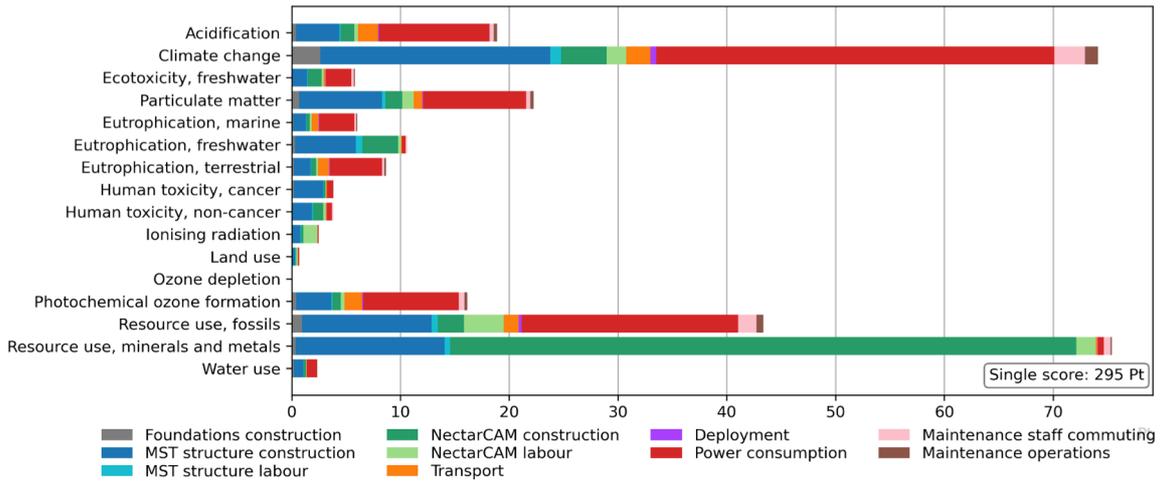

**Figure 3. Weighted life cycle impacts of one MSTN.** The values reflect integrated impacts from construction and 30 years of operations. Weighted values are expressed in units of Pt, corresponding to the annual environmental load of one average person in the world.

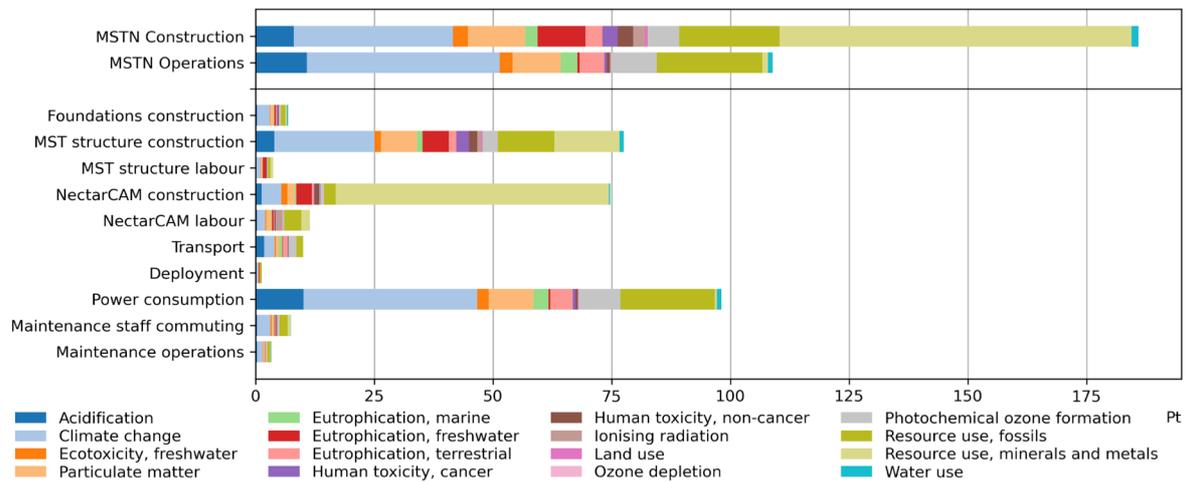

**Figure 4. Single score life cycle impacts of one MSTN.** Scores, expressed in units of Pt, correspond to the annual environmental load of one average person in the world. Impacts in the top section are shown for construction and operations, where we integrated the deployment impacts into the bar for construction. Impacts in the bottom section are shown by activity.

# Discussion

Our analysis reveals that both construction and operations contribute substantially to the environmental impacts of the telescope. This confirms earlier findings by Knödlseder et al. [8] who showed that both contributions are important for the life cycle carbon footprint of ground-based observatories. Our study also confirms the order of magnitude of the carbon footprint of observatory construction that was estimated in that study, and that was related to the construction costs through a monetary emission factor. Knödlseder et al. [8] found a value of 240 kgCO$_2$eq / k€ for observatory construction, while dividing the 932 tCO$_2$eq of emissions that we found for MSTN construction by an estimated construction cost of 3 M€ leads to a slightly larger value of 311 kg CO$_2$eq / k€ (see 'CO$_2$ equivalent emission factors' in Supplementary Information). The difference may be explained by the different scope of the emission factors, as observatory construction includes not only telescopes, but also items such as buildings, roads or computing infrastructures that were not considered in our study.

In the simplified LCA for the GRANDProto300 project, Vargas-Ibáñez et al. [10] found that mineral, metal and fossil resource use as well as climate change were the dominant impact categories for telescope construction, and we come to the same conclusion for the life cycle impacts of MSTN. Interestingly, the same impact categories dominate also for the development of space instruments [12] with metal and mineral resources use impacts being also related to the manufacturing of instrument hardware. On the other hand, fossil resources use and climate change impacts in the space sector are originating primarily from instrument testing in clean rooms as well as office work and travelling, while for the MSTN they originate primarily from telescope construction and power consumption during operations.

Concerning telescope construction, metal casting turns out to contribute considerably to the environmental impacts of telescope construction, suggesting improvements if the casting processes could be replaced by less impacting manufacturing processes, for example by using machined pieces with simpler geometries. We note that metal casting was chosen to reduce the investment cost for MST-STR construction, hence avoiding this process will probably increase the cost of the telescope structure. If the casting process for the head and yokes of the telescope positioner were replaced by steel metal working, which is used for the other parts of the telescope, the environmental life cycle impacts of MSTN would be reduced between 5% and 25%, depending on impact category. Specifically, the climate change impact of one MSTN would be lowered by 381 tCO$_2$eq, corresponding to 14% of the life cycle impact.

Avoiding use of steel where possible presents another avenue for impact reductions for telescope construction. For example, were the steel-made counterweights replaced by concrete blocks of the same weight, environmental life cycle impacts of MSTN would be reduced in all categories, reaching up to 13% for human cancer toxicity, while the climate change impact would be reduced by 89 tCO$_2$eq, corresponding to 3% of the MSTN life cycle impact. Yet concrete blocks would take a larger volume, needing a redesign of the structure that eventually may offset some of the impact reductions.

Many environmental impacts from construction are also associated with mining activities and the treatment of generated waste. Avenues for impact reductions include therefore the reduction of the amounts of material used, the use of recycled materials, and the tracing of materials through the supply chain with the goal to avoid landfills for waste management. This applies in particular to the camera, where the use of minerals and metals (specifically gold) for integrated circuits used in the electronics provides a major contribution to the environmental impacts. Here, design changes that minimise the amount of integrated circuits needed in the camera may provide the largest potential for impact reductions, since reuse and recycling of these electronics are in many cases still not feasible. While it may be too late for such design changes in the initial camera, they should definitely be pursued when considering the camera upgrade, to be expected after 15 years of operations.

Furthermore, environmental impacts of many manufacturing processes are driven by the carbon intensity of the electricity mix used for production. We assumed that most MST-STR manufacturing and assembly is carried out in Germany, where fossil fuels such as coal and natural gas contribute substantially to electricity production. Selecting companies that can demonstrate the use of renewable energies for manufacturing (beyond claims from green certificates, see Brander et al. [23]), or shifting manufacturing to countries with a low carbon intensity of the electricity mix, are options to be explored for impact reductions.

Reducing the carbon intensity of energy-use in the buildings where assembly, integration and verification activities are carried out, summarised in our study by the FTE category, also provides a way to reduce environmental impacts, in particular for the categories of ionising radiation, land use and use of fossil resources. This includes powering office buildings by renewable energies as well as thermal insulation of buildings to reduce needs for heating and air conditioning. The environmental impact of labour can also be reduced by acting on home-to-office commuting and professional travelling, for example by favouring alternatives to commuting by car and replacing in person meetings by video conferencing solutions [7].

The largest environmental impact of the operations phase originates from the electricity consumption for operating the telescope and camera on La Palma, assuming that the electricity will continue to be provided by the current energy system of La Palma that heavily relies on diesel generators. To reduce this impact, a first step would be the minimisation of the energy consumption of the MSTN system. About 50% of the energy is consumed in the parking or standby states, during which no science data is acquired (see 'Energy consumption during operations' in Supplementary Information). It should be investigated how this power consumption can be minimised, for example by switching off systems that are not relevant for the survival of the instruments. Furthermore, it should be investigated whether the consumption of principal consumers, such as the Front End Boards (FEBs) of the camera, the camera cooler, or the drive assemblies of the structure, can be reduced.

Complementary to reducing the consumption, the use of renewable energy sources should be investigated on La Palma, with potential co-benefits for all other astronomical observatories on ORM, and the island in general. Currently, only four wind turbines are installed on La Palma, which produced for example 23 GWh of electricity in the year 2015 [24]. Comparing this amount to the 53 MWh of electricity needed to operate one MSTN for one year shows that a single wind turbine, associated to some energy storage system [25], would be amply sufficient to power all 13 CTAO telescopes that will be installed on La Palma. Assuming that all diesel generators on La Palma would be replaced by a wind turbine powered hydro storage system alike the one currently operating on the neighboring island El Hierro would reduce the carbon footprint of electricity consumption during operations from 1,362 $tCO_2eq$ to 106 $tCO_2eq$. The total MSTN life cycle carbon footprint would be reduced from 2,660 $tCO_2eq$ to 1,449 $tCO_2eq$ by this single measure, presenting a 46% reduction in GHG emissions (see also Extended Data Fig. 3). It is obvious that using renewable energies, and in particular wind turbines, which are highly efficient on La Palma, in combination with a hydro storage system for energy storage, provides the most dramatic reduction of the MSTN environmental footprint that can be achieved without changing the performance or scope of the system.

The MSTN construction on La Palma is now imminent and given the advanced state of the project it is unlikely that the environmental impacts of construction can be significantly reduced. The focus has hence to be put on reducing the environmental impacts during operations. The first aspect to be considered is the minimisation of the energy consumption for the telescope structure and camera, eventually by considering less power-consuming electronics or motors that may go at the expense of slightly degraded scientific performance. The second aspect to envision is the implementation of a renewable energy system at ORM. Detailed plans of how to implement such a system that also includes energy storage to assure continuous energy supply are still to be developed (see, e.g., Viole et al. [11]), and co-benefits with other optical observatories at ORM and the local island communities are to be explored.

Our study makes clear that environmental impacts need to be considered upfront in a research infrastructure project, well before it comes to the stage of implementation. We propose to do this by adding environmental impact budgets to the project requirements, making sure that they are treated on equal footing with scientific performance and technical specifications. LCAs should be performed early on, probably in a simplified manner during the early stages of the project, and then progressively refined as the project definition improves. In that way, the outcomes of the LCA can inform the project specifications and guide the manufacturing towards the least impacting processes. LCA results should also inform implementation decisions, eventually resulting in the abandonment of projects that do not fit into environmental impact budgets. Scientific research will only become sustainable once environmental limits are respected, which is a prerequisite to stabilise the Earth system and bring it back into the planetary boundaries.

# Methods

**Method, tools and assumptions.** We followed the ISO 14040 standard [15] for our assessment, using the SimaPro software version 9.5.0 in combination with the LCI database ecoinvent version 3.9.1. In case where a suitable process was not available in ecoinvent, a customised process was modelled. We divided the inventories into Electronics, Mechanics, Manufacturing, Transport, Electricity and Full-Time Equivalent (FTE) categories for better visualisation and discussion of results. We used the Environmental Footprint 3.1 (adapted) method version 1.00 for the impact analysis, and applied a cut-off approach for allocation.

For processes where the origin of the product was unknown, market processes were used as they include different origins that are weighted according to market share. For assembly processes we explicitly took the transportation between the place of origin and assembly of the components into account. Regarding the geographical location of processes, the following order of priority was followed. We preferred country-specific processes when the origin was known and when it was available as a choice in the ecoinvent database. European (RER) average processes were used as the second-best option, since except for the camera support structure of the telescopes that is provided by a research institute located in Brazil, all other elements are provided by European research institutes. Global (GLO) and Rest of World (RoW) processes were used when the first two options were not available.

**Boundaries of the system.** The system boundaries are depicted in Extended Data Fig. 1, and an overview of the MSTN is provided in Supplementary Figure 1. Our assessment is limited to the construction, on-site deployment and operations of one MST on the CTAO northern site in La Palma, excluding the early definition and prototyping as well as decommissioning processes and activities.

The early definition and prototyping activities were spread over many years, and the exploratory nature of this phase combined with the lack of detailed bookkeeping prevents including this phase in the assessment. Plans for the decommissioning phase were not available in time for the study, and consequently this phase was excluded from the analysis. We note, however, that astronomical facilities are rarely dismantled, and often used for educational or outreach purposes once they are no longer scientifically useful. Even dismantled telescopes are often not thrown away, but elements are stored, and possibly reused later, as illustrated for example by the FACT Cherenkov Telescope that is a refurbished version of the HEGRA telescope [16]. Hence excluding the end-of-life phase corresponds to a plausible scenario for MSTN.

**Organisation of data collection and uncertainties.** We gathered inventory data, such as mass, material, quantity and dimensions, from technical documents provided by members of the MST-STR and NectarCAM consortia, complemented by data obtained through email exchanges with consortia members, direct measurements, and own estimates. All data and their references were collected in excel spreadsheets including Bills Of Materials (BOM) and corresponding LCI processes that were identified in the ecoinvent database. For each entry in the spreadsheet we determined an uncertainty related to the data accuracy and an uncertainty related to the correspondence with a process of the ecoinvent database. We combined both uncertainties in a confidence matrix that we matched to the ecoinvent pedigree matrix. In that way, we determined for each LCI entry a single uncertainty factor for a log-normal uncertainty distribution that we list together with the flow data in the excel spreadsheets. Details on the uncertainty computation are provided in 'Handling of uncertainties' in the Supplementary Information. Based on these uncertainty factors, we determined the 95% confidence level uncertainties for each impact category using Monte Carlo simulations performed with SimaPro. The file collecting all spreadsheets with the information used for our assessment is available for download at https://zenodo.org/records/11915488.

**Consideration of labour.** Life cycle analyses of space systems reveal that office work, and labour in general, contribute significantly to the environmental impacts of the construction of a scientific satellite due to the energy consumption of the buildings where the work is executed [17]. To understand whether the same is true for ground-based observatories we also assessed labour needed for construction and maintenance, as measured by the number of FTE needed to fulfil a task. Following the prescription of the European Space Agency (ESA) handbook for life cycle assessments [18], we based the characterisation factors of the environmental impact of labour on the inventory for running the office buildings, purchase of laptops, and home-to-office commuting of personal of the Institut de Recherche en Astrophysique et Planétologie (IRAP) determined in the context of the carbon footprint estimate of the institute in 2019 [19]. This resulted in GHG emissions of 2.8 tCO$_2$eq per FTE for IRAP, and we adopt the corresponding characterisation factors for labour in France. By adapting the underlying ecoinvent processes based on location we derived an equivalent characterisation factor for labour in Germany, resulting in 5.8 tCO$_2$eq per FTE. For La Palma we developed a specific model that excludes daily commuting (since commuting was modelled explicitly, see below), resulting in 7.7 tCO$_2$eq per FTE.

**Inventory for foundation production.** We separated the inventory for the foundation from that of the telescope structure, since the responsibility for providing the former resides with the CTAO while the structure is provided by a scientific consortium led by DESY. The LCI data for the foundations, together with their correspondence in the database, are summarised in Supplementary Table 1. The masses of the foundations, made of reinforced concrete, were estimated from the drawings.

**Inventory for MST-STR production.** We established the inventory for the telescope structure production by decomposing the system into its major parts and their components. The resulting LCI data are provided in Supplementary Table 2.

Most elements are primarily made of steel and aluminium, for which masses and production processes were extracted from project documentation. For all metal parts employed in the structure, as well as in the camera, we separated the metal production processes, listed under the category "Mechanics", from the metal transformation processes, listed under the category "Manufacturing", including casting, rolling, coating and general metal working. Specifically, casting processes, which turned out to have significant environmental impacts due to the involved energy consumption (see below), are employed for the manufacturing of the head and yokes of the steel-made positioner, and the aluminium-made Mirror Support Units (MSUs). Further manufacturing processes include the fabrication of plastic parts, the coating of the mirrors, and the production of honeycomb structures, which were modelled by the aluminium sheet rolling process.

The telescope mirrors are components that were specifically developed for CTAO telescopes, composed of hexagonal sandwiched mirror segments that are produced using the glass cold-slumping replication technology [20]. We modelled the production of these segments by a mechanical structure made of flat glass and aluminium honeycomb, connected using epoxy resin, and coated by a mix of aluminium, silicon dioxide and zirconium dioxide, using densities of 2,500 kg/m$^3$ for flat glass, 1,250 kg/m$^3$ for epoxy, and 104 kg/m$^3$ for the honeycomb.

The telescope structure also comprises electronic parts, including motors for the azimuth and elevation drives as well as the Active Mirror Control (AMC) system, electronic boards (modelled as lead free surface mounted printed wiring boards), cables, power supply units, batteries, computers and routers. Specifically, the AMC actuators were divided into aluminium-made mechanics, electronic boards and a motor. The mass of the electronic boards was estimated from mass/area ratios of other boards used in the telescope, the mass of the motor was taken from the manufacturer, and the remaining mass was attributed to mechanics. For the pointing camera, which is part of the calibration system, we divided the mass of the system into a camera (modelled as an unspecified active electronic component), an embedded computer, and an aluminium-made housing.

Finally, the telescope structure comprises auxiliary devices including cabinets, Uninterruptible Power Supply (UPS), a cabling system, a chiller for camera cooling and humidity control, ladders, lightning protection and additional counter weights. Following expert advice, we modelled the UPS as a combination of router and lithium batteries, while the chiller was modelled as a combination of Acrylonitrile-butadiene-styrene (ABS) copolymer for the mechanical parts, and a heat pump plus a power supply unit for the electronic parts.

Transport includes both shipping of components to the integration site of DESY, located in Zeuthen, Germany, and the shipping of telescope elements from the production and integration sites to the CTAO north site ORM on La Palma Spain. All transport distances were estimated using the Routescanner online tool (https://www.routescanner.com). Most telescope structure components are produced by industry with delivery to DESY included in the production and manufacturing processes. Delivery of pre-integrated components from DESY to ORM is modelled as a mix of road and sea transport, with lorries of Euro 3 class > 32 metric tons used for the former. The Camera Support Structure (CSS) is provided by a research institute located in São Paulo, Brazil, and shipped directly to ORM via a mix of road and sea transport, modelled using lorries of Euro 3 class 3.5 - 7.5 metric tons used for the former. Mirrors are provided by the Media Lario company located in Boisio Parini, Italy, and shipped directly to ORM via a mix of road and sea transport applying the same model as for the CSS. For shipping of telescope components, component masses were multiplied by a factor 1.5 to consider the weight of any packaging and protection material, as recommended by [18]. Finally, we also modelled the shipping of rebars required for the foundations from Madrid, Spain, to ORM, via a mix of Euro 3 class 3.5 - 7.5 metric tons lorries and sea freight.

The labour in the research institutes needed to produce, pre-assemble and validate one telescope structure is estimated to be 6 FTE, which is modelled as German labour since most of the activities will take place in Germany.

**Inventory for NectarCAM production.** Similarly to the telescope structure, we established the inventory for the production of the NectarCAM camera by decomposing the system into its major parts and their components. The resulting LCI data, together with their correspondence in the database, are summarised in Supplementary Table 3.

The camera modules are composed of Focal Plane Modules (FPM), Front End Boards (FEB), and backplanes, for which masses of all components were determined by direct measurement. All electronic boards were modelled as surface mounted lead-free printed wiring boards, with explicit modelling of some heavy integrated circuits (see below). FPMs comprise 7 Detector Units (DU), each composed of a Winston cone that serves as light concentrator, a Photomultiplier Tube (PMT), a High Voltage and Pre-Amplification board (HVPA) comprising an Application-Specific Integrated Circuit (ASIC), and some mechanical elements. The DUs are connected to an Interface Board (IB) and integrated into a mechanical structure. The Winston cone was modelled as an Acrylonitrile-Butadiene-Styrene (ABS) copolymer, excluding the reflective layer due to lack of information, yet with an expected negligible impact due to its thinness. The ASIC on the HVPA board was explicitly modelled as an integrated circuit of memory type. The PMT is a complex component composed of borosilicate glass coated by mu-metal (iron-nickel soft magnetic material) and an insulating cover, a super bialkali photocathode composed of antimony, potassium and caesium, and a dynode structure composed of copper and beryllium. We determined a mass of 48 g for the PMT without coatings, and split this mass between internal electronics, glass tube, and coatings on the basis of the dimensions and density of the materials. By analogy, the internal electronics were then modelled as an electron gun for cathode ray tube displays, while borosilicate glass was used for the tube, nickel-rich material for the mu-metal, and ethyl vinyl acetate foil for the insulating cover. Each FEB is composed of a printed circuit board and an aluminium frame with a total weight of 404 g. ASICs and the Nectar Field Programmable Grid Array (FPGA) were modelled separately from the remaining board as integrated circuits.

Regarding the mechanical structure of the camera, the most commonly used material is aluminium, followed by plastic and some small steel parts. For the camera housing we developed a customised model since the material used is honeycomb that is not available in ecoinvent. We modelled the panels as a mix of polyurethane foam, epoxy resin and glass fibre reinforced plastic polyester resin with material amounts based on dimensions that we extracted from the technical documentation and densities of 40 kg/m$^3$, 1,250 kg/m$^3$ and 1,650 kg/m$^3$, respectively. The camera front is composed of an acrylic window that we modelled as a polymethyl methacrylate sheet, and a shutter that we modelled as polyester fibre.

The cooling system is constituted of a heat exchanger, fans, a cooling fluid and the mechanical structure. For the heat exchanger the total mass was divided between copper, steel and aluminium in the proportions of 60%, 25% and 15% that we estimated from product photos, an estimate that we did not refine due to its small contribution with respect to the other impacts of the camera. Fans were modelled as products for desktop power supply units, and ethylene glycol was used to model the cooling fluid.

The power system was modelled as power supplies for desktop computers, a model that we did not refine due to its small contribution. The digital trigger was modelled as a combination of an aluminium-made mechanical structure and lead-free surface mounted printed wiring board, and the data acquisition by an internet router. Unspecified cables and an aluminium-made mechanical structure were used to model the cable management system, and a mix of internet routers, desktop computers and aluminium-made mechanical structure was used to model the calibration devices. The sensor system was modelled by assuming that two thirds of the mass was copper and the rest passive unspecified electronic components, cable trays were considered to be made of aluminium.

Transport includes shipping of components from supplying research institutes located in France, Spain and Germany to the camera integration facility at IRFU, Saclay, France, where shipping is modelled using Euro 3 class lorries of 3.5 - 7.5 metric tons capacity, multiplying the shipped weights again by a factor 1.5 to consider protection and packaging. Following assembly and testing, the camera is shipped from Saclay, France, to the ORM site via a combination of road transport and sea freight. The labour in the research institutes needed to produce, assemble and validate one NectarCAM is estimated to be 11 FTE, which is modelled as French labour since the camera is essentially developed and built in France.

**Inventory for construction phase.** Based on the inventories for telescope structure and cameras we compiled the total inventory for the construction phase of one MST telescope for the CTAO north site. This includes one foundation, one MST-STR constructed for the entire life cycle of CTAO and the construction of two NectarCAMs. We assumed the construction of two NectarCAMs since after 15 years of operations it is expected that the camera will be upgraded by replacing either parts or the entire camera by a new system; since no plans or specifications exist so far for the second generation camera, we assumed that its environmental impacts will be identical to those of the NectarCAM. In addition, spare parts are produced for structure and camera to replace items that will fail during the operations phase. Based on a Reliability, Availability, Maintainability and Safety (RAMS) analysis of the system, it is assumed that an average of 5% of spare parts are needed for the structure and the cameras. Hence the inventory of the construction phase is computed using

$$LCI(construction) = LCI(Foundation) + 1.05 \times LCI(MST\text{-}STR) + 2.1 \times LCI(NectarCAM)$$

taking into account all hardware that will be produced for operating one MST during the CTAO life cycle of 30 years. Note that transport activities of all elements to the ORM site in La Palma are included in the inventory of the construction phase. The resulting LCI is shown in Supplementary Table 4.

**Inventory for on-site deployment.** Once all elements arrive on the ORM site in La Palma they need to be assembled, integrated and commissioned. We estimated the labour needed for these activities based on an internal report. We also considered the transport of scientists, engineers and technicians to the site, assuming air-travelling from Barcelona to Santa Cruz de La Palma, as well as daily transport to the site carried out during the 38 days of assembly and 16 days of commissioning. The resulting LCI is shown in Supplementary Table 5.

**Inventory for operations phase.** During the operations phase, environmental impacts are generated by the telescope through the consumption of energy and through maintenance activities.

Energy consumption was modelled using a customised process that reflects the current annual electricity production mix on the island, with 89.5% of the power generated using diesel, 8.3% from wind, 2.0% from photovoltaic power plants and 0.2% from natural gas [21]. As energy consumption of the telescope structure and the camera varies according to operating state, we estimated the annual duration in each state, and multiplied those durations by the estimated average power consumption in each state as specified in the technical documentation or determined by expert judgement. The results are summarised in Supplementary Table 7. In total, the annual energy consumption amounts to 53.0 MWh of which 22.6 MWh are for the structure and 30.4 MWh for the camera. Alternative scenarios based on an increased usage of renewable energy on La Palma are explored in 'Alternative energy systems for La Palma' in the Supplementary Information.

For maintenance, we accounted for the commuting of maintenance staff to ORM, assuming a rota scheme of 8 days on site followed by 6 days off. The total annual maintenance activities for one MSTN are estimated to 312 person-hours per year, while the maintenance of all CTAO telescopes on ORM is estimated to 4,690 person-hours per year. This means that maintenance can be covered by two persons being permanently on site, which implies with the assumed rota scheme 85 annual return flights, of which 312/4,690 are attributed to the maintenance of one MSTN. Using the same attribution, we also assume that two persons commute daily by car from Santa Cruz to ORM. The resulting LCI data for the operations of one MSTN are summarised in Supplementary Table 6.

# Data Availability

All data used for this study are available for download at https://zenodo.org/records/11915488

# Acknowledgements

We would like to thank all members of the MSTN and NectarCAM collaborations who supported this study by providing information and data. We are grateful to Markus Garczarczyk and Oscar Ferreira for having granted permission to use their drawings in Supplementary Figure 1. We also thank Xavier Loizillon from the Scalian company for valuable discussions. We furthermore thank the Centre National d'Etudes Spatiales (CNES) for providing financial support that allowed the acquisition of licences of the SimaPro software and ecoinvent databases, as well as a dedicated laptop that was used during this study. Finally, we thank the members of the Labos1point5 research group for valuable discussions and for providing training on the Life Cycle Assessment method. This study was financed in part by the Coordenação de Aperfeiçoamento de Pessoal de Nível Superior - Brasil

# Author Contributions Statement

G.d.S.I. conducted the LCA and provided a detailed report on which this article is based. She also contributed to the writing of the article. M.B., J.K., P.G., and L.M. designed the study, supervised the LCA, and contributed to the writing of the article. P.J., G.P., A.S. and F.T. served as technical experts for the LCA, provided information to establish the LCI, and reviewed the article.

# Correspondence

Correspondence should be adressed to J.K. (jurgen.knodlseder@irap.omp.eu).

# Competing Interests

The authors declare no competing interests.

# References

[1] Richardson, K., Steffen, W., Lucht, W., et al., *Earth beyond six of nine planetary boundaries*, Sci. Adv. **9**, eadh2458 (2023)

[2] IPCC *Summary for Policymakers. In: Climate Change 2021: The Physical Science Basis. Contribution of Working Group I to the Sixth Assessment Report of the International Panel on Climate Change*, Cambridge University Press (2021)

[3] IPBES *Global assessment report of the Intergovernmental Science-Policy Platform on Biodiversity and Ecosystem Services*, Brondizio, E. S., Settele, J., Díaz, S., Ngo, H. T. (eds.). IPBES secretariat, Bonn, Germany. ISBN: 978-3-947851-20-1 (2019)

[4] IRP *Global Resources Outlook 2019: Natural Resources for the Future We Want.* Oberle, B., Bringezu, S., Hatfield-Dodds, S., et al. (eds.) A Report of the International Resources Panel. United Nations Environment Programme. Nairobi, Kenya (2019)

[5] Baste, I. A., Watson, R. T. *Tackling the climate, biodiversity and pollution emergencies by making peace with nature 50 years after the Stockholm Conference*, Global Environmental Change **73**, 102466 (2022)

[6] Rosen, J. *A greener culture*, Nature **546**, 565-567 (2017)

[7] Knödlseder, J. *in: Climate Change for Astronomers: Causes, consequences, and communication*, Rector, T. (ed.), AAS-IOP, ISBN: 978-0750337250 (2024)

[8] Knödlseder, J., Brau-Nogué, S., Coriat, M., et al. *Estimate of the carbon footprint of astronomical research infrastructures*, Nat. Astron. **6**, 503-513 (2022)

[9] Kruithof, G., Bassa, C., Bonati, I., et al. *The energy consumption and carbon footprint of the LOFAR telescope*, Exp. Astron. **56**, 687-714 (2023)

[10] Vargas-Ibáñez, L. T., Kotera, K., Blanchard, O., et al. *Life Cycle Analysis of the GRAND Experiment*, Astropart. Phys. **155**, 102903 (2024)

[11] Viole, I., Shen, L., Camargo, L. R., et al. *Sustainable Astronomy: A comparative life cycle assessment of off-grid hybrid energy systems to supply large telescopes*, Int J Life Cycle Assess (2024)

[12] Barret, D., Albouys, V., Knödlseder, J., et al. *Life Cycle Analysis of the Athena X-ray Integral Field Unit*, Exp. Astron. **57**, 19 (2024)

[13] Hofmann, W., Zanin, R. *The Cherenkov Telescope Array* in: Handbook of X-ray and Gamma-ray Astrophysics, (Eds. C. Bambi and A. Santangelo), Springer, Singapore. https://doi.org/10.1007/978-981-16-4544-0_70-1 (2024)

[14] Cabrera, S. M. *La Palma como espejo*, Energías Renovables, 11 December. https://www.energias-renovables.com/panorama/la-palma-como-espejo--20231211 (2023)

[15] International Organization for Standardization. *Environmental management - Life cycle assessment - Principles and framework. ISO 14040:2006-10*, ICS: 13.020.60 (2022)

[16] Bretz, T., Anderhub, H., Backes, M., et al. *A status report: FACT – a fact!* in: Proc. 13th ICATPP, (Eds. G. Simone, et al.), World Scientific Publishing Co. Pte. Ltd., 29-33 (2012)

[17] Chanoine, A., Duvernois, P.-A., Le Guern, Y. *Environmental impact assessment analysis. Technical note D8*, https://nebula.esa.int/sites/default/files/neb_study/1116/C4000104787ExS.pdf, accessed on 8 November 2023 (2017)

[18] ESA LCA Working Group *Space system Life Cycle Assessment (LCA) guidelines*, ESSB-HB-U-005, Issue 1, Revision 0 (2016)

[19] Martin, P., Brau-Nogué, S., Coriat, M., et al. *A comprehensive assessment of the carbon footprint of an astronomical institute*, Nat. Astron. **6**, 1219-1222 (2022)

[20] La Palombara, N., Sironi, G., Giro, E., et al. *Mirror production for the Cherenkov telescopes of the ASTRI mini-array and the MST project for the Cherenkov Telescope Array*, J. Astron. Tel. Inst. Sys. **8**, 1, 014005 (2022)

[21] Anuario Energético de Canarias, https://www3.gobiernodecanarias.org/ceic/energia/oecan/files/AnuarioEnergeticoCanarias_2021_v2.pdf, accessed on 9 November 2023 (2021)

[22] Sala, S., Cerutti, A. K., Pant, R. *Development of a weighting approach for the Environmental Footprint.* Joint Research Centre (JRC) 106546 (2018)

[23] Brander, M., Gillenwater, M., Ascui, F. *Creative accounting: A critical perspective on the market-based method for reporting purchased electricity (scope 2) emissions*. Energy Policy **112**, 29-33 (2018)

[24] Sistema Eléctrico Español: Avance (2015) REE (Red Electrica de España). https://www.ree.es/sites/default/files/downloadable/avance_informe_sistema_electrico_2015_v2.pdf

[25] Ramirez-Diaz, A., Ramos-Real, F. J., Marrero, G. A. *Complementarity of electric vehicles and pumped-hydro as energy storage in small isolated energy systems: case of La Palma, Canary Islands*. J. Mod. Power Syst. Clean Energy **4**, 604 - 614 (2016)

[26] Hunter, J. D. *Matplotlib: A 2D Graphics Environment.* Computing in Science & Engineering **9**, 90-95 (2007).

# Extended Data Figures

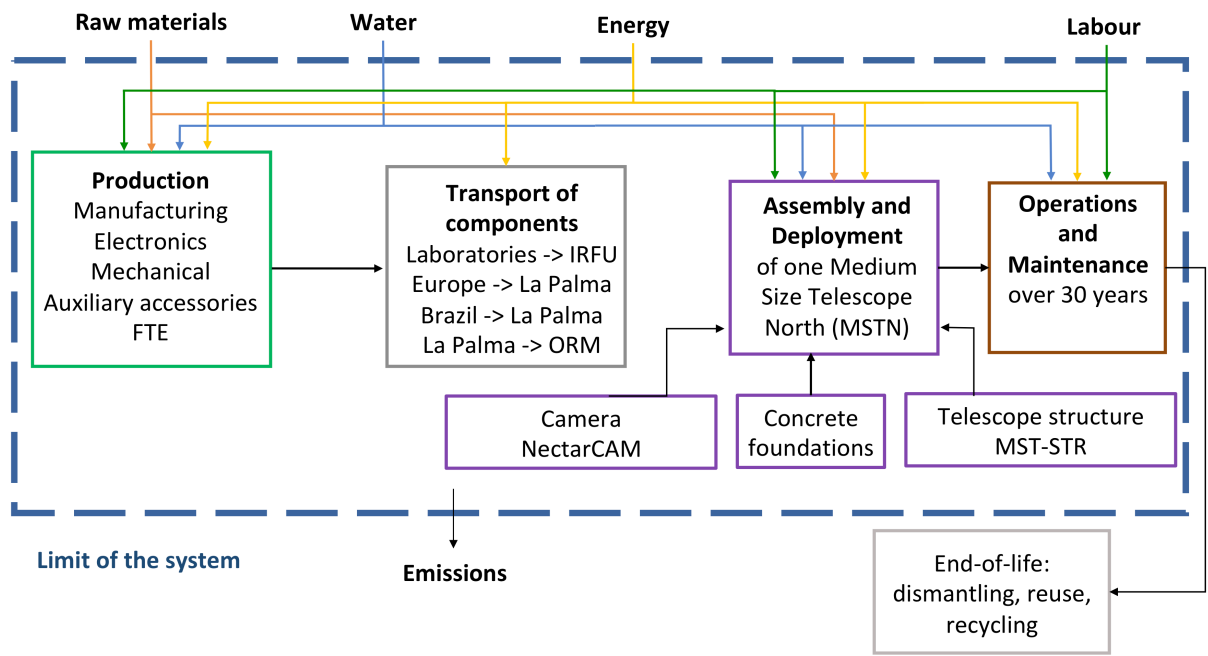

**Extended Data Figure 1. System boundary of the life cyle assessment.** The end-of-life impacts were excluded from the analysis as decommissioning plans for MSTN are not yet elaborated.

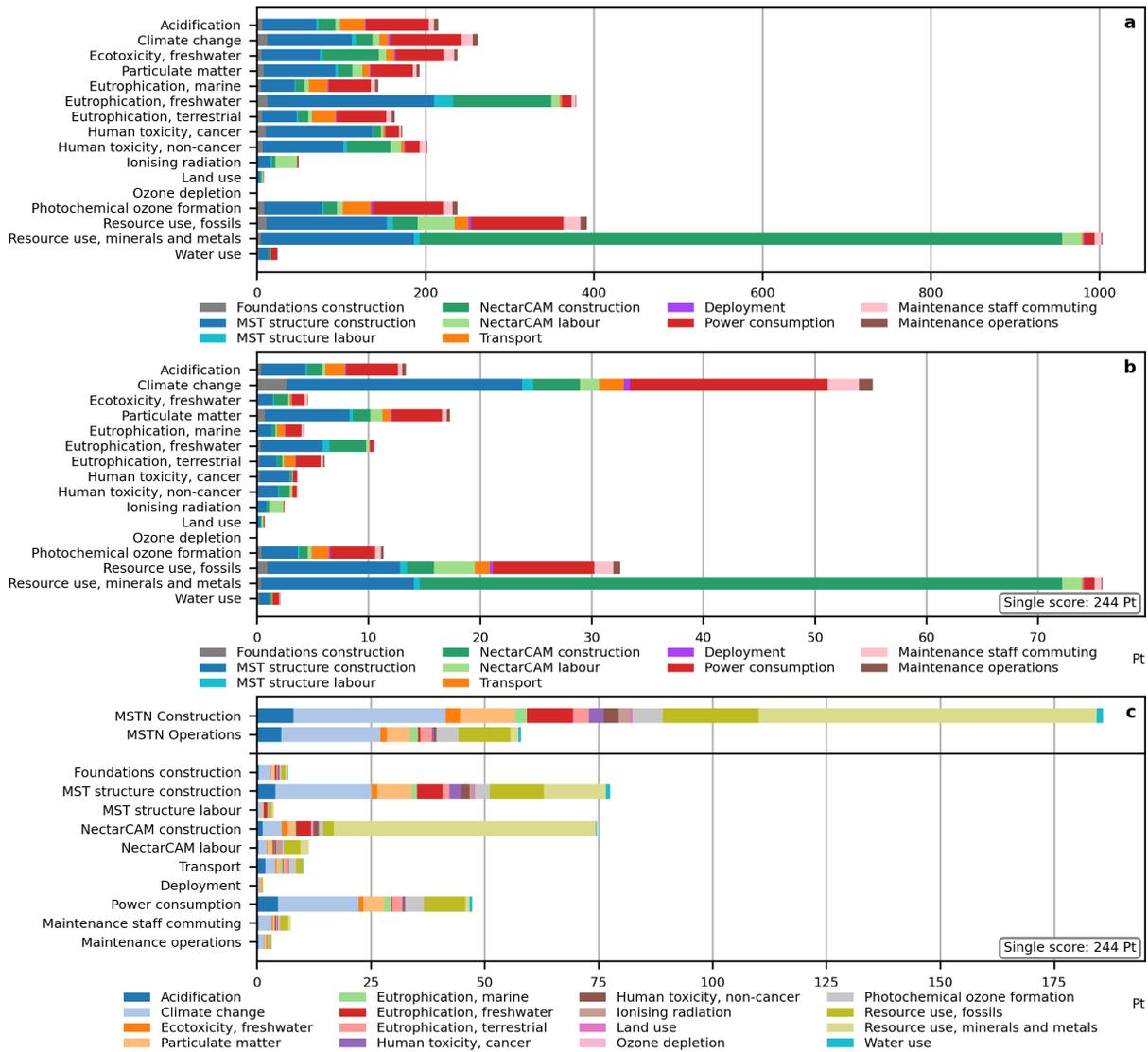

**Extended Data Figure 2. Normalised (a), weighted (b) and single score (c) life cycle impacts of one MSTN for an alternative energy system on La Palma.** The results correspond to 30 years of operations with an alternative energy system, assuming that a wind powered pumped hydro storage system is implemented in La Palma like the one that exists on the island of El Hierro. We assumed that 56% of the electricity demand that is currently covered by diesel generators on La Palma would be replaced by electricity from this system, which corresponds to the contribution that is currently achieved in El Hierro (see Supplementary Information for details).

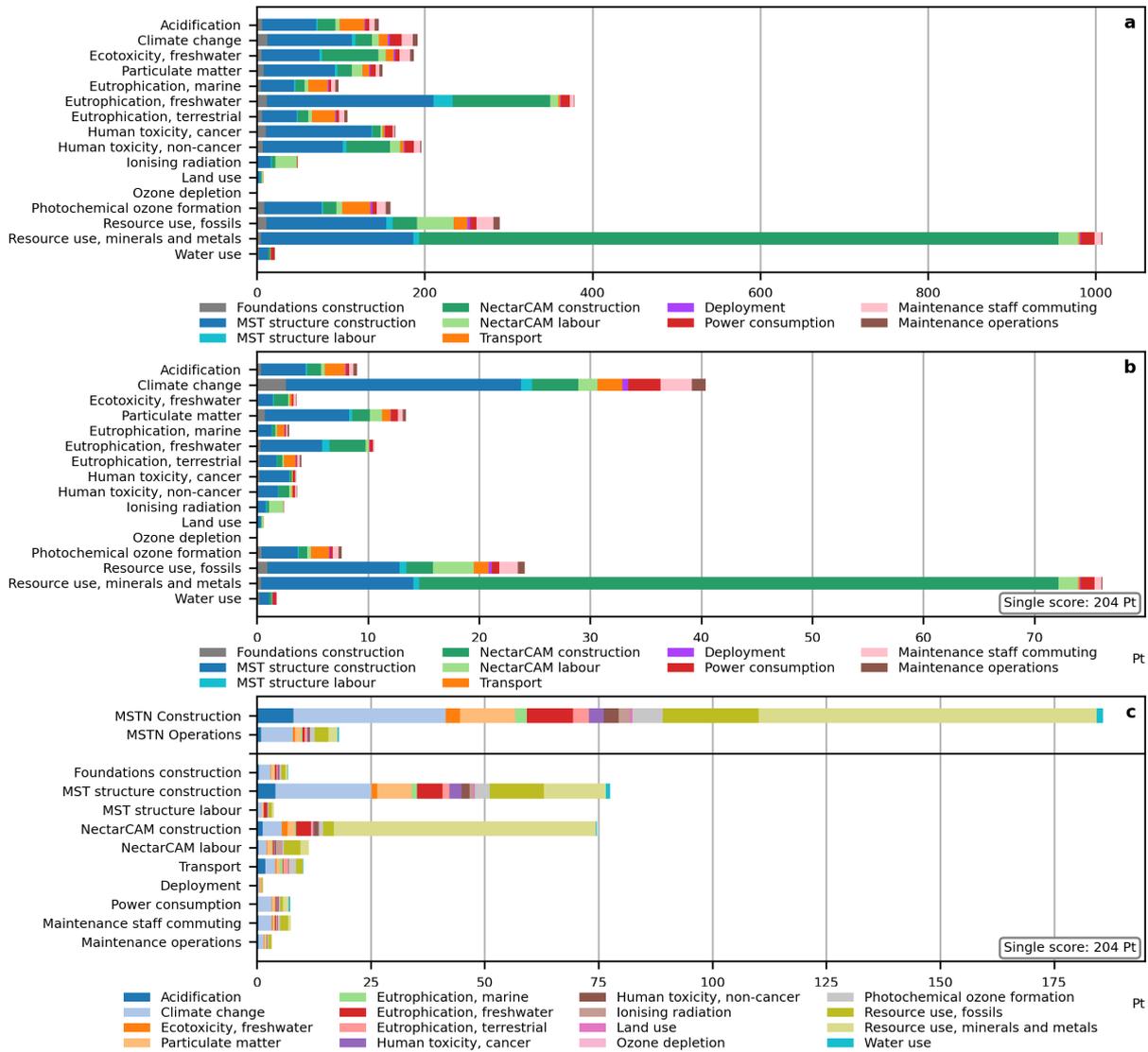

**Extended Data Figure 3. Normalised (a), weighted (b) and single score (c) life cycle impacts of one MSTN for an alternative energy system on La Palma.** The results correspond to 30 years of operations with an alternative energy system, assuming that a wind powered pumped hydro storage system is implemented in La Palma like the one that exists on the island of El Hierro. We assumed that 100% of the electricity demand that is currently covered by diesel generators on La Palma would be replaced by electricity from this system (see Supplementary Information for details).

# Supplementary Information

## Overview over the Mid-Sized Telescope

The system that we studied is a Mid-Sized Telescope (MST) deployed on the Cherenkov Telescope Array Observatory (CTAO) northern site (MSTN) in La Palma, Spain. The MST consists of a telescope structure (MST-STR), provided by an international consortium of research institutes led by the Deutsches Elektronen-Synchrotron (DESY), located in Zeuthen, Germany, and a NectarCAM camera, provided by an international consortium of research institutes led by the Institut de Recherche sur les lois Fondamentales de l'Univers (IRFU), located in Saclay, France.

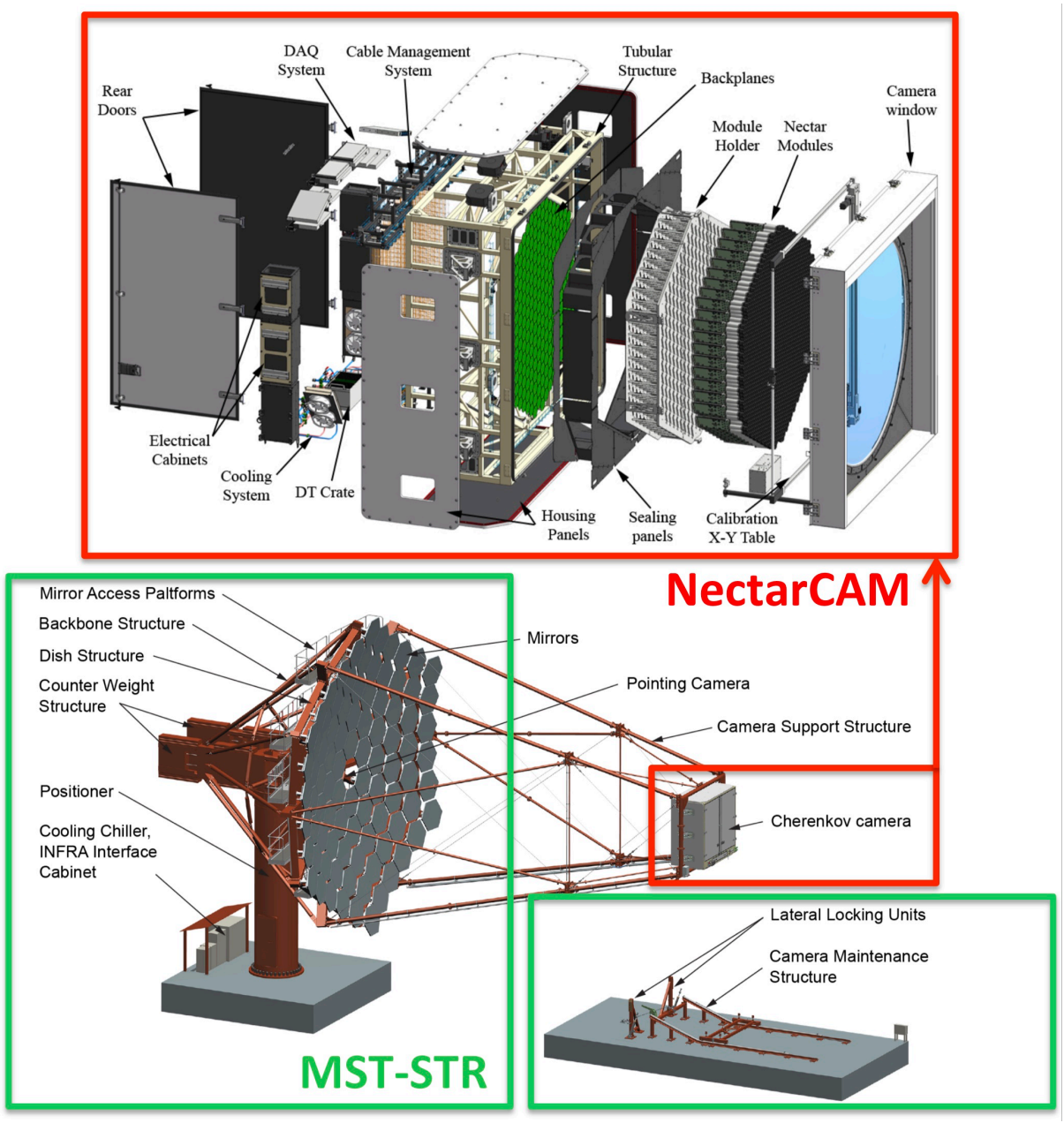

Supplementary Figure 1. Representation of the MST system, comprising MST-STR and NectarCAM.

Supplementary Figure 1 depicts the MST system and its elements. The MST has a maximum height of 28.5 m pointing vertically and 15.9 m pointing horizontally, a width of 25.8 m and a dish diameter of 13.8 m. The NectarCAM has a height of 2.75 m and a width of 1.7 m. The MST-STR weighs about 82 t and the NectarCAM 2.2 t. The MST-STR is anchored at a tower foundation that consists of 106.9 m$^3$ of concrete, with length and width of 7 m and a depth of 2.2 m. When the telescope is not in use, the camera is stored on a camera maintenance structure foundation, consisting of 81.68 m$^3$ of concrete, with length of 13.7 m, width of 6 m and a depth of 1 m.

We modelled the MST-STR and NectarCAM units independently, including an estimated amount of 5% of spare parts that will be used for maintenance during the 30 years of operations. After 15 years of operations, it is expected that the camera will be upgraded by replacing either parts or the entire camera by a new system. Since no plans or specifications exist so far for the second generation camera, we assumed that its environmental impacts will be identical to those of the NectarCAM. The life cycle therefore comprises the construction of one MST-STR and two NectarCAMs, and operations of a system composed of MST-STR and NectarCAM during 30 years. Operations include electricity consumed for operating the telescope, as well as labour and associated travelling required for the maintenance of the system.

## Life cycle inventory data

The following tables summarise the life cycle inventory (LCI) data collected for this study. Quantities quoted in tons refer to metric tons.

**Supplementary Table 1. LCI data for the production of one telescope foundation.**

| Category | Description | Unit | Quantity |
|---|---|---|---|
| Mechanics | Concrete, 35 Mpa {RoW} | market | m³ | 189 |
| Mechanics | Reinforcing steel {GLO} |market | ton | 11.2 |
| Mechanics | Steel, unalloyed {GLO} | market | kg | 594 |
| Manufacturing | Metal working, average for steel product manufacturing {RER} | kg | 594 |

**Supplementary Table 2. LCI data for the production of one MST-STR.**

| Category | Description | Unit | Quantity |
|---|---|---|---|
| Electronics | Battery cell, Li-ion, LFP {GLO} | market | kg | 43 |
| Electronics | Cable, unspecified {GLO} | market | kg | 370 |
| Electronics | Computer, desktop, without screen {GLO} | market | p | 2 |
| Electronics | Electric motor, for electric scooter {GLO} | market | kg | 402 |
| Electronics | Electronic component, active, unspecified {GLO} | market | g | 700 |
| Electronics | Heat pump, brine-water, 10 kW {RoW} | p | 1 |
| Electronics | Power supply unit, for desktop computer {GLO} | market | p | 27 |
| Electronics | Printed wiring board, surface mounted, unspecified, Pb free {GLO} | market | kg | 8.6 |
| Electronics | Router, internet {GLO} | market | p | 49 |
| Mechanics | Acrylonitrile-butadiene-styrene (ABS) copolymer {GLO} | market | kg | 131 |
| Mechanics | Aluminium alloy, AlMg3 {GLO} | market | ton | 3.2 |
| Mechanics | Aluminium, cast alloy {GLO} | market | kg | 324 |
| Mechanics | Aluminium oxide, non-metallurgical {IAI Area, EU27 & EFTA} | market | kg | 18.9 |
| Mechanics | Copper-rich materials {GLO} | market | kg | 40 |
| Mechanics | Epoxy resin insulator, SiO2 {GLO} | market | kg | 12.1 |

**Supplementary Table 2 continued**

| Category | Description | Unit | Quantity |
|---|---|---|---|
| Mechanics | Epoxy resin, liquid {RER} | market | kg | 47.7 |
| Mechanics | Flat glass, coated {RER} | market | ton | 1 |
| Mechanics | Steel, unalloyed {GLO} | market | kg | 74.9 |
| Mechanics | Steel, chromium steel 18/8 {GLO} | market | kg | 1.07 |
| Mechanics | Zirconium oxide {GLO} | market | kg | 21.7 |
| Manufacturing | Casting, aluminium, lost-wax {RoW} | kg | 324 |
| Manufacturing | Casting, steel, lost-wax {RoW} | ton | 11 |
| Manufacturing | Metal working, average for aluminium product manufacturing {RER} | ton | 2.51 |
| Manufacturing | Metal working, average for aluminium product manufacturing {RoW} | kg | 388 |
| Manufacturing | Metal working, average for chromium steel product manufacturing {RER} | ton | 1.07 |
| Manufacturing | Metal working, average for copper product manufacturing {RER} | kg | 40 |
| Manufacturing | Metal working, average for steel product manufacturing {RER} | ton | 60.1 |
| Manufacturing | Metal working, average for steel product manufacturing {RoW} | ton | 3.81 |
| Manufacturing | Powder coat, aluminium sheet {RER} | m² | 95.5 |
| Manufacturing | Sheet rolling, aluminium {RER} | kg | 298 |
| Transport | Transport, freight, lorry 3.5-7.5 metric ton, EURO 3 {RER} | market | ktkm | 1.6 |
| Transport | Transport, freight, lorry 3.5-7.5 metric ton, EURO 3 {BR} | market | ktkm | 1.97 |
| Transport | Transport, freight, sea, ferry {GLO} | market | ktkm | 378 |
| Transport | Transport, freight, lorry >32 metric ton, EURO 3 {RER} | market | ktkm | 218 |
| Transport | Transport, freight, lorry 7.5-16 metric ton, EURO 3 {RER} | market | ktkm | 6.28 |
| FTE | FTE {DE} | FTE | 6 |

**Supplementary Table 3. LCI data for the production of one NectarCAM.**

| Category | Description | Unit | Quantity |
|---|---|---|---|
| Electronics | Cable, unspecified {GLO} | market | kg | 104 |
| Electronics | Computer, desktop, without screen {GLO} | market | p | 1 |
| Electronics | Electronic component, passive, unspecified {GLO} | market | kg | 1 |
| Electronics | Electron gun, for cathode ray tube display {GLO} | market | kg | 38.5 |
| Electronics | Fan, for power supply unit, desktop computer {GLO} | market | kg | 27.3 |
| Electronics | Integrated circuit, logic type {GLO} | market | g | 795 |
| Electronics | Integrated circuit, memory type {GLO} | market | kg | 1.06 |
| Electronics | Light emitting diode {GLO} | market | kg | 3.2 |
| Electronics | Power supply unit, for desktop computer {GLO} | market | p | 4 |
| Electronics | Printed wiring board, surface mounted, unspecified, Pb free {GLO} | market | kg | 158 |
| Electronics | Router, internet {GLO} | market | p | 9 |
| Mechanics | Acrylonitrile-butadiene-styrene (ABS) copolymer {GLO} | market | kg | 70.2 |
| Mechanics | Aluminium alloy, AlMg3 {GLO} | market | ton | 1.09 |
| Mechanics | Aluminium, wrought alloy {GLO} | market | kg | 97.3 |
| Mechanics | Copper-rich materials {GLO} | market | kg | 16 |
| Mechanics | Epoxy resin, liquid {RER} | market | kg | 27.2 |
| Mechanics | Ethylene glycol {GLO} | market | kg | 15.5 |
| Mechanics | Ethylvinylacetate, foil {GLO} | market | kg | 3.71 |
| Mechanics | Fibre, polyester {GLO} | market | kg | 40 |

**Supplementary Table 3 continued**

| Category | Description | Unit | Quantity |
|---|---|---|---|
| Mechanics | Glass tube, borosilicate {GLO} \| market | kg | 50.6 |
| Mechanics | Glass fibre reinforced plastic, polyester resin, hand lay-up {GLO} \| market | kg | 182 |
| Mechanics | Nickel-rich materials {GLO} \| market | kg | 28 |
| Mechanics | Polymethyl methacrylate, sheet {GLO} \| market | kg | 32 |
| Mechanics | Polyurethane, flexible foam {RER} \| market | kg | 20.9 |
| Mechanics | Steel, chromium steel 18/8 {GLO} \| market | kg | 63.1 |
| Mechanics | Steel, unalloyed {GLO} \| market | kg | 25 |
| Manufacturing | Extrusion, co-extrusion {FR} | kg | 40 |
| Manufacturing | Extrusion of plastic sheets and thermoforming, inline \| market | kg | 228 |
| Manufacturing | Injection moulding {RER} | kg | 102 |
| Manufacturing | Metal working, average for aluminium product manufacturing {RER} | ton | 1.19 |
| Manufacturing | Metal working, average for chromium steel product manufacturing {RER} | kg | 63.1 |
| Manufacturing | Metal working, average for copper product manufacturing {RER} | kg | 16 |
| Manufacturing | Metal working, average for steel product manufacturing {RER} | kg | 25 |
| Transport | Transport, freight, lorry 3.5-7.5 metric ton, EURO 3 {RER} \| market | ktkm | 4.18 |
| Transport | Transport, freight, sea, ferry {GLO} \| market | ktkm | 7.6 |
| FTE | FTE {FR} | FTE | 11 |

**Supplementary Table 4. LCI data for the construction phase of one MSTN.**

| Category | Description | Unit | Quantity |
|---|---|---|---|
| Electronics | Battery cell, Li-ion, LFP {GLO} \| market | kg | 45.1 |
| Electronics | Cable, unspecified {GLO} \| market | kg | 607 |
| Electronics | Computer, desktop, without screen {GLO} \| market | p | 4.2 |
| Electronics | Electric motor, for electric scooter {GLO} \| market | kg | 423 |
| Electronics | Electronic component, active, unspecified {GLO} \| market | g | 735 |
| Electronics | Electronic component, passive, unspecified {GLO} \| market | kg | 2.1 |
| Electronics | Electron gun, for cathode ray tube display {GLO} \| market | kg | 80.8 |
| Electronics | Fan, for power supply unit, desktop computer {GLO} \| market | kg | 57.4 |
| Electronics | Heat pump, brine-water, 10 kW {RoW} | p | 1.05 |
| Electronics | Integrated circuit, logic type {GLO} \| market | kg | 1.67 |
| Electronics | Integrated circuit, memory type {GLO} \| market | kg | 2.23 |
| Electronics | Light emitting diode {GLO} \| market | kg | 6.72 |
| Electronics | Power supply unit, for desktop computer {GLO} \| market | p | 36.8 |
| Electronics | Printed wiring board, surface mounted, unspecified, Pb free {GLO} \| market | kg | 342 |
| Electronics | Router, internet {GLO} \| market | p | 70.3 |
| Mechanics | Acrylonitrile-butadiene-styrene (ABS) copolymer {GLO} \| market | kg | 285 |
| Mechanics | Aluminium alloy, AlMg3 {GLO} \| market | ton | 5.65 |
| Mechanics | Aluminium, cast alloy {GLO} \| market | kg | 340 |
| Mechanics | Aluminium oxide, non-metallurgical {IAI Area, EU27 & EFTA} \| market | kg | 19.9 |
| Mechanics | Aluminium, wrought alloy {GLO} \| market | kg | 204 |
| Mechanics | Concrete, 35 MPa {RoW} \| market | m³ | 189 |
| Mechanics | Copper-rich materials {GLO} \| market | kg | 75.5 |
| Mechanics | Epoxy resin, liquid {RER} \| market | kg | 107 |

**Supplementary Table 4 continued**

| Category | Description | Unit | Quantity |
|---|---|---|---|
| Mechanics | Epoxy resin insulator, SiO2 {GLO} | market | kg | 12.7 |
| Mechanics | Ethylene glycol {GLO} | market | kg | 32.6 |
| Mechanics | Ethylvinylacetate, foil {GLO} | market | kg | 7.79 |
| Mechanics | Fibre, polyester {GLO} | market | kg | 84 |
| Mechanics | Flat glass, coated {RER} | market | ton | 1.05 |
| Mechanics | Glass tube, borosilicate {GLO} | market | kg | 106 |
| Mechanics | Glass fibre reinforced plastic, polyester resin, hand lay-up {GLO} | market | kg | 383 |
| Mechanics | Nickel-rich materials {GLO} | market | kg | 58.8 |
| Mechanics | Polymethyl methacrylate, sheet {GLO} | market | kg | 67.2 |
| Mechanics | Polyurethane, flexible foam {RER} | market | kg | 43.8 |
| Mechanics | Reinforcing steel {GLO} | market | ton | 11.2 |
| Mechanics | Steel, chromium steel 18/8 {GLO} | market | ton | 1.26 |
| Mechanics | Steel, unalloyed {GLO} | market | ton | 79.3 |
| Mechanics | Zirconium oxide {GLO} | market | kg | 23 |
| Manufacturing | Casting, aluminium, lost-wax {RoW} | kg | 340 |
| Manufacturing | Casting, steel, lost-wax {RoW} | ton | 11.6 |
| Manufacturing | Extrusion, co-extrusion {FR} | kg | 84 |
| Manufacturing | Extrusion of plastic sheets and thermoforming, inline | market | kg | 478 |
| Manufacturing | Injection moulding {RER} | kg | 215 |
| Manufacturing | Metal working, average for aluminium product manufacturing {RER} | ton | 5.15 |
| Manufacturing | Metal working, average for aluminium product manufacturing {RoW} | kg | 407 |
| Manufacturing | Metal working, average for chromium steel product manufacturing {RER} | ton | 1.26 |
| Manufacturing | Metal working, average for copper product manufacturing {RER} | kg | 75.5 |
| Manufacturing | Metal working, average for steel product manufacturing {RER} | ton | 63.8 |
| Manufacturing | Metal working, average for steel product manufacturing {RoW} | ton | 4.01 |
| Manufacturing | Powder coat, aluminium sheet {RER} | m² | 100 |
| Manufacturing | Sheet rolling, aluminium {RER} | kg | 313 |
| Transport | Transport, freight, lorry 3.5-7.5 metric ton, EURO 3 {RER} | market | ktkm | 10.5 |
| Transport | Transport, freight, lorry 3.5-7.5 metric ton, EURO 3 {BR} | market | ktkm | 2.07 |
| Transport | Transport, freight, lorry 7.5-16 metric ton, EURO 3 {RER} | market | ktkm | 6.6 |
| Transport | Transport, freight, lorry >32 metric ton, EURO 3 {RER} | market | ktkm | 228 |
| Transport | Transport, freight, sea, ferry {GLO} | market | ktkm | 413 |
| FTE | FTE {FR} | FTE | 23.1 |
| FTE | FTE {DE} | FTE | 6.3 |

**Supplementary Table 5. LCI data for the deployment phase of one MSTN.**

| Category | Description | Unit | Quantity |
|---|---|---|---|
| Transport | Transport, passenger aircraft, medium haul {GLO} | market | pkm | 53,086 |
| Transport | Transport, passenger car, medium size, petrol, EURO 3 {RER} | km | 9,343 |
| FTE | FTE {La Palma} | FTE | 1.14 |

**Supplementary Table 6. LCI data for the operations phase of one MSTN.**

| Category | Description | Unit | Quantity |
|---|---|---|---|
| Electricity | Electricity, high voltage {ES} | electricity production, oil | TJ | 5.12 |
| Electricity | Electricity, high voltage {ES} | electricity production, natural gas, combined cycle power plant | GJ | 5.72 |
| Electricity | Electricity, high voltage {ES} | electricity production, natural gas, conventional power plant | GJ | 5.72 |
| Electricity | Electricity, high voltage {ES} | electricity production, wind, 1-3 MW turbine, onshore | GJ | 475 |
| Electricity | Electricity, low voltage {ES} | electricity production, photovoltaic, roof | GJ | 57.2 |
| Electricity | Electricity, low voltage {ES} | electricity production, photovoltaic, ground | GJ | 57.2 |
| Transport | Transport, passenger aircraft, medium haul {GLO} | market | pkm | 440,370 |
| Transport | Transport, passenger car, medium size, petrol, EURO 3 {RER} | km | 133,736 |
| FTE | FTE {La Palma} | FTE | 5.71 |

# Mass balance

In order to verify completeness of the inventories we computed the masses of the telescope structure and camera according to our LCIs. For the telescope foundation, concrete volumes were converted into weights assuming a density of 2400 kg/m$^3$. Items specified by amount unit (p) instead of mass unit (kg), such as the masses of computers, power supplies, routers and heat pumps were excluded. The results for the structure are summarised in Supplementary Table 7, separated into mass with and without telescope foundations. In total, our inventory reflects a total mass of 81.6 metric tons for the telescope structure and 464.4 metric tons for the foundation. The results for the camera are summarised in Supplementary Table 8. Our inventory reflects a total mass of 2,100 kg of which 84% are attributed to mechanics and 16% to electronics. This compares to a mass of 2,193 kg as specified in the technical documents, where the difference of 4% is plausibly explained by the items excluded from the mass balance.

**Supplementary Table 7. Summary of the mass of one foundation and one MST-STR for each category in the inventory.**

| Category | Description | Mass with Foundations (kg) | | Mass without Foundations (kg) | |
|---|---|---|---|---|---|
| Foundation | Concrete and steel | 464,387 | 85.05% | - | - |
| Mechanics | Steel; aluminium; plastic; glass... | 80,808 | 14.80% | 80,808 | 98.99% |
| Electronics | Boards; components; cables... | 824 | 0.15% | 824 | 1.01% |
| | **Total** | **546,019** | | **81,632** | |

**Supplementary Table 8. Summary of the mass of one NectarCAM for each category in the inventory.**

| Category | Description | Mass (kg) | |
|---|---|---|---|
| Mechanics | Steel; aluminium; plastic; glass... | 1,765.21 | 84% |
| Electronics | Boards; components; cables; computers; routers... | 334.49 | 16% |
| | **Total** | **2,099.67** | |

# Handling of uncertainties

To handle uncertainties, we implemented a confidence matrix for the LCI data collection to trace the data quality and the correspondence of identified ecoinvent database entries. As shown in Supplementary Table 9, the accuracy of flow data and the correspondence with the database entry was classified on a scale from 0 to 4, with 0 indicating the absence of data or database entry, and 4 indicating a direct measurement and a perfect database match. The matrix provides insights into which processes are well-modelled in our analysis, and which processes may warrant improvements in a next iteration of the assessment.

**Supplementary Table 9. Confidence matrix used to evaluate the quality of input data and the correspondence with the ecoinvent database.**

| Flow data accuracy | | Correspondence with database entry | |
|---|---|---|---|
| Level | Meaning | Level | Meaning |
| 0 | No data provided | 0 | No entry found that can be remotely linked to the flow data |
| 1 | Unreliable data, rough estimate, remote proxy | 1 | The entry is a proxy that could be representative (estimate) |
| 2 | Proxy, estimates | 2 | The entry is a proxy from the same process/material family |
| 3 | Reliable data based on expert assumption or literature | 3 | Close or same process/material |
| 4 | Direct measurement | 4 | Perfect match |

We matched the levels to the uncertainty factors of the ecoinvent pedigree matrix to assign for each combination of levels in the confidence matrix an uncertainty factor. The pedigree approach is widely used in LCA and estimates quantitative uncertainties based on qualitative characteristics of a data set [1]. The uncertainty factors (UFs) that we derived from the matching of our confidence matrix to the ecoinvent pedigree matrix are given in Supplementary Table 10. In addition, we used a base uncertainty factor of 1.05 for all energy-related, material and manufacturing processes, and a base uncertainty factor of 2 for all transport services that were added to the uncertainty factors of Supplementary Table 10 to derive total uncertainty factors using

$$\text{Total UF} = \exp(\sqrt{\ln(\text{Base UF})^2 + \ln(\text{Supplementary Table 10 UF})^2})$$

The uncertainties of all MSTN LCI data were then modelled in SimaPro as lognormal distributions.

**Supplementary Table 10. Uncertainty factors for the confidence matrix.** Flow accuracy levels are given as rows, database entry levels are given in columns.

| Flow \ Database | 0 | 1 | 2 | 3 | 4 |
|---|---|---|---|---|---|
| 0 | 2.68 | 2.25 | 2.07 | 2.03 | 2.02 |
| 1 | 2.25 | 1.80 | 1.58 | 1.54 | 1.53 |
| 2 | 2.07 | 1.58 | 1.33 | 1.27 | 1.24 |
| 3 | 2.02 | 1.53 | 1.25 | 1.17 | 1.14 |
| 4 | 2.02 | 1.53 | 1.24 | 1.16 | 1.12 |

# Energy consumption during operations

Energy consumption of the telescope structure and the camera during operations varies according to operating state, hence to derive the total energy consumption we estimated the annual duration in each state, and multiplied those durations by the estimated average power consumption in each state as specified in the technical documentation or determined by expert judgement. The results are summarised in Supplementary Table 11. In total, the annual energy consumption amounts to 53.0 MWh of which 22.6 MWh are for the structure and 30.4 MWh for the camera.

**Supplementary Table 11. Power needs and estimated annual energy consumption per operating state.**

| State | Annual duration | MST-STR | | NectarCAM | |
|---|---|---|---|---|---|
| | (hours/year) | Power (W) | Energy (kWh) | Power (W) | Energy (kWh) |
| Parking | 5,608 | 2,100 | 11,777 | 1,759 | 9,864 |
| Standby | 1,020 | 3,200 | 3,264 | 1,759 | 1,794 |
| Tracking | 2,102 | 3,300 | 6,937 | 8,799 | 18,495 |
| Repositioning | 30 | 20,000 | 600 | 8,799 | 264 |
| **Total annual** | **8,760** | | **22,577** | | **30,418** |

# Alternative energy systems for La Palma

We have assumed in our modelling that the electricity system on La Palma will stay as it currently is over next 30 years, with 89.5% of electricity demand satisfied by diesel generators, 0.2% by gas turbines, 8.3% by wind turbines, and 2.0% by solar panels [2]. This assumption may be conservative in view of the stated goal by the clean energy for EU island secretariat of the European Commission to make La Palma an island powered to 100% by renewable energies in 2030 [3], but we are not aware of any concrete implementation plans for such a renewable energy system, and it is questionnable whether the goal will be actually reached [3].

The neighbouring island of El Hierro is the most advanced of the Canary islands in the transition towards renewable energy systems, being equipped with an innovative wind turbine powered pumped hydro storage system that currently covers 56% of the energy demand of the island, the remaining demand being almost entirely covered by diesel generators [4]. While the energy demand on El Hierro is about 20% of that on La Palma [2], one may optimistically assume that a similar (yet about five times larger) energy system is implemented rapidly on La Palma, replacing 56% of the electricity demand that is currently satisfied by diesel generators with a wind turbine powered pumped hydro storage system. That would lead to an electricity mix on La Palma where 39.4% of the demand is satisfied by diesel generators, 50.3% by the wind turbine powered pumped hydro storage system, 8.3% by the already existing wind turbines, and 2.0% by already existing solar panels. To model such a scenario, we created a customised process for the wind turbine powered pumped hydro storage system, which we based on the "hydro, pumped storage" process for Spain, replacing the Spanish electricity mix by that of wind turbines (1-3 MW, on-shore) in Spain. This reduces the lifecycle carbon footprint of one MSTN by 26% from an initial value of 2,660 ± 274 $tCO_2eq$ to 1,979 ± 196 $tCO_2eq$. The resulting environmental life cycle impacts of one MSTN in this scenario are shown in Extended Data Fig. 2.

One may even push this analogy one step further, and assume that 100% of the current diesel generation on La Palma is replaced by a wind turbine powered pumped hydro storage system, which needs plausibly to be about ten times larger than the system that exists on El Hierro, involving the installation of about 50 wind turbines that are similar in size to those on El Hierro (2.3 MW). This would reduce the lifecycle carbon footprint of one MSTN from an initial value of 2,660 ± 274 $tCO_2eq$ to 1,449 ± 180 $tCO_2eq$, which corresponds to a 46% reduction. The resulting environmental life cycle impacts of one MSTN in this scenario are shown in Extended Data Fig. 3.

## $CO_2$ equivalent emission factors

$CO_2$ equivalent ($CO_2eq$) emission factors have been used in the literature to estimate the carbon footprint of research infrastructures in the past. Specifically, Knödlseder et al. [5] determined a monetary emission factor of 240 $kgCO_2eq$ / k€ for the construction of ground-based astronomical observatories from the literature. Assuming the cost for structure and camera to roughly 1.5 M€ each, and based on the masses of the subsystems, we give in Supplementary Table 12 monetary and mass emission factors for telescope foundation, structure, camera and the MSTN without foundations. In particular we derive a monetary emission factor of 311 kg $CO_2eq$ / k€ for the latter, which is a bit larger than the value derived by Knödlseder et al. [5] for observatory construction. Obviously, a complete observatory comprises more than a single telescope, combining items with a variety of emission factors, which may explain the different values.

**Supplementary Table 12. Emission factors for $CO_2eq$ emission from construction activities.**

| Item | $kgCO_2eq$ | Mass (kg) | Cost (k€) | $kgCO_2eq$ / kg | $kgCO_2eq$ / k€ |
|---|---|---|---|---|---|
| Telescope foundation | 94,402 | 464,387 | | 0.2 | |
| Telescope structure | 827,023 | 81,632 | 1,500 | 10 | 551 |
| Camera | 105,178 | 2,100 | 1,500 | 50 | 70 |
| MSTN w/o foundation | 932,201 | 83,732 | 3,000 | 11 | 311 |

# References

[1] Qin, Y., Cucurachi, S., Suh, S. *Perceived uncertainties of characterization in LCA: a survey* Int. J. Life Cycle Assess. **25**, 1846 - 1858 (2020)

[2] Anuario Energético de Canarias, https://www3.gobiernodecanarias.org/ceic/energia/oecan/files/AnuarioEnergeticoCanarias_2021_v2.pdf, accessed on 9 November 2023 (2021)

[3] Cabrera, S. M. La Palma como espejo, Energías Renovables, 11 December. https://www.energias-renovables.com/panorama/la-palma-como-espejo--20231211 (2023)

[4] Gioda, A. El Hierro (Canaries) : une île et le choix des transitions énergétique et écologique, VertigO, 14 (2014)

[5] Knödlseder, J., Brau-Nogué, S., Coriat, M. et al. *Estimate of the carbon footprint of astronomical research infrastructures*, Nat. Astron. **6**, 503-513 (2022)